\documentclass[twocolumn, english,aps,prl,reprint, floatfix,notitlepage,footinbib,nosuperscriptaddress]{revtex4-2}
\usepackage{lmodern}

\usepackage[T1]{fontenc}
\usepackage[latin9]{inputenc}
\usepackage{geometry}
\usepackage{subfigure,lmodern, amsmath,amssymb, graphicx, pifont, adjustbox, bm, xcolor}
\usepackage{amsfonts}
\usepackage{comment}
\usepackage{mathtools}
\usepackage{float}
\usepackage{slashed}
\usepackage{ragged2e}
\usepackage{array}
\usepackage{tensor}
\usepackage{tcolorbox}  
\usepackage{stackengine}
\newcommand{\appendixref}[1]{\hyperref[#1]{appendix~\ref{#1}}}
\def\equationautorefname~#1\null{eq.\,(#1)\null}
\usepackage[hang,flushmargin]{footmisc} 
\usepackage{cleveref}

\allowdisplaybreaks
\makeatletter

\usepackage{etoolbox}
\apptocmd{\thebibliography}{\justifying\setlength{\leftskip}{7.4mm}}{}{} 
 
\usepackage{relsize}
\usepackage{babel}
\usepackage{changepage}

\newcommand{\be}{\begin{equation}}
\newcommand{\ee}{\end{equation}}
\newcommand{\bea}{\begin{eqnarray}}
\newcommand{\eea}{\end{eqnarray}}

\DeclareMathOperator\arccosh{arccosh}

\begin{document}

\title{Conservative Dynamics of Relativistic Binaries Beyond Einstein Gravity}

\author{Jordan Wilson-Gerow}
\affiliation{Department of Physics, Carnegie Mellon University, Pittsburgh, PA 15213, USA}

\begin{abstract}
\noindent 

We study the conservative dynamics of spinless compact objects in a general effective theory of gravity which includes a metric and an arbitrary number of scalar fields, through $\mathcal{O}(G^{3})$. Departures from Einstein gravity, which preserve general coordinate and local Lorentz invariance, are characterized by higher-derivative terms in a Lagrangian whose coupling constants scale as powers of a ``new-physics'' length scale, $\ell$.  For a purely metric theory we compute the contributions from the the leading and subleading higher-curvature curvature corrections. In four dimensions these are cubic and quartic curvature terms, i.e. orders $\ell^4$ and $\ell^6$.  We also study a general multi-scalar-tensor theory of gravity to order  $\ell^{4}$, which includes both Einstein-dilaton-Gauss-Bonnet and dynamical Chern-Simons  higher-order couplings. Specifically, we compute the radial action in a post-Minkowskian approximation for scattering orbits, to two-loop order. The result encodes the fully relativistic dynamics of the compact objects, and serves as a generating function for gauge-invariant orbital observables for both bound and unbound binary systems. Where overlapping post-Newtonian results are available, we've verified agreement.

\end{abstract}

\preprint{}

\maketitle

\noindent {\bf Introduction:} Strong-field gravity is now a precision science. Pulsar timing arrays offer a view of the persistent unruly background of gravitation radiation permeating our universe~\cite{Agazie_2023}, and the event horizon telescope gives a direct image of churning of plasma in the nearest possible region of supermassive black holes~\cite{2019ApJ...875L...1E}. These landmark observations are, however, observations of complicated stochastic systems. In contrast, laser interferometric gravitation wave detectors allow us to carefully listen to the inspiral and merger of relatively isolated compact astrophysical bodies. This comparatively clean environment allows for the experimental precision to be matched by commensurate theoretical precision. By comparing observational signals with Numerical Relativity waveform templates, the LIGO-Virgo-KAGRA (LVK) collaboration has been able to precisely measure properties of black holes and neutron stars~\cite{PhysRevX.13.041039}. These measurements provide unparalleled information into astrophysical systems~\cite{Buonanno:2014aza}.

These observations also present an opportunity to test fundamental physics, in particular, to test the limit to which gravity is indeed described by General Relativity (GR)~\cite{PhysRevD.94.084002}. To do so one needs a parameterized deformation of the signatures predicted by GR, so that GR can be treated as the null hypothesis. Such deformations can be broadly classified as model-independent and model-dependent. In a model-dependent test one computes the same observable in a specific non-GR theory. A primary limitation here is the computational cost required to cover a sufficiently broad class of models. In a model-independent test (e.g.~\cite{PhysRevD.80.122003}) one introduces a general deformation of the observable, but there is, in principle, no a priori knowledge determining: \textit{ i)} which regions of the parameter space are consistent with physical principles, and \textit{ii)} if the parameter space is sufficiently general to cover all models of interest. For a comprehensive review of tests of GR using both approaches, see~\cite{Yagi:2016jml}\,.

Effective Field Theory (EFT) offers a resolution to these issues when there is a hierarchy of physical scales. Subject to a given set of physical principles, it is a tool for computing the most general corrections to an observable, at least perturbatively in the ratio of these physical scales. 

There is a large literature studying models which modify GR (see e.g. \cite{Fujii:2003pa, Sotiriou:2008rp, ALEXANDER20091, DeFelice:2010aj, Faraoni:2010pgm, CLIFTON20121, Will:2018bme}). These models generically involve higher curvature modifications to the Einstein-Hilbert Lagrangian, and/or adding scalar degrees of freedom as in Jordan-Fierz-Brans-Dicke theory~\cite{Fierz:1956zz, Jordan:1959eg, PhysRev.124.925} and its generalizations to Damour-Esposito-Farese tensor multiscalar gravity~\cite{Damour:1992we}. With tests of GR as a goal, binary dynamics in scalar-tensor models in particular,  have been well studied in a post-Newtonian approximation~\cite{Damour:1992we, PhysRevD.53.5541, PhysRevD.54.1474, PhysRevD.87.084070, PhysRevD.89.084014, PhysRevD.91.084027, PhysRevD.94.084003, Julie:2019sab, Bernard:2022noq, Bernard:2023eul, Almeida:2024uph}, including the state of the art 3PN results for spinless bodies~\cite{Bernard:2018hta, Bernard:2018ivi, Julie:2022qux}. 

It is impossible to identify, let alone to compute observables in, all possible models which modify GR. Nonetheless if we impose the physical principles of locality, general coordinate invariance, and local Lorentz invariance, we can use EFT to make completely general predictions for binary dynamics, with which any models respecting these principles must agree---at least perturbatively in $\ell/b$, where $\ell$ is the ``new-physics'' length scale and $b$ is the orbital scale. The idea of General Relativity as an EFT is far from new (see e.g.~\cite{Donoghue:1994dn, Burgess:2003jk}), and there is a long history of computing black hole solutions which account for string-theory inspired corrections,~\cite{Callan:1988hs, Campbell:1990ai, Campbell:1990fu, PhysRevD.43.3140, Campbell:1991kz, Campbell:1991rz, Campbell:1992hc, Shapere:1991ta, Mignemi:1992nt, Mignemi:1993ce}, as well as studying them for observational signatures~\cite{PhysRevLett.99.241101, PhysRevD.75.124022, Moura:2006pz, PhysRevD.79.084031, ALEXANDER20091, Yunes:2009hc,Yunes:2011we, Yagi:2012ya, PhysRevD.85.102003, Stein:2014xba}.

In the context of gravitational wave science, the EFT of gravity has been discussed in \cite{Endlich:2017tqa, Cardoso:2018ptl, deRham:2020ejn}, where black hole solutions, quasinormal modes, and the post-Newtonian approximation were studied. In that work the only low energy degree of freedom was the metric. However, many models of interest also include light scalar fields which are presumed to interact with Standard Model fields only with gravitational strength. To accommodate this, in this work we will generalize the previous EFT to allow for an arbitrary number of scalar fields and their higher-derivative coupling to gravity.

With the effective theory in hand, one still needs to compute observables. Computational costs for numerical relativity currently make scanning theory-space unfeasible, especially while the parameter space of GR itself (e.g. spin magnitudes and alignment angles) has not yet been fully scanned. Analytic computations do not require scanning parameter space, and are more amenable to generic EFT studies. For the binary problem these primarily come in two perturbative approaches: post-Newtonian (PN) or the fully relativistic post-Minkowskian (PM). 

The recent import of modern scattering amplitudes methods and insights from EFT (see e.g.~\cite{Smirnov:2012gma, dixon2013-review, ElvangHuang, cheung2017tasi, Neill:2013wsa, Kalin:2020mvi} and ~\cite{Cheung:2018wkq, Bjerrum-Bohr:2018xdl, Bjerrum-Bohr:2022blt}), have led to a boom in PM results. Landmark work at $\mathcal{O}(G^3)$ (3PM) ~\cite{PhysRevLett.122.201603, Bern:2019crd} rapidly lead to many results, notably the conservative and dissipative dynamics at 4PM order~\cite{Bern:2021dqo, Bern:2021yeh, Dlapa:2021vgp,  Dlapa:2022lmu, Damgaard:2023ttc, jakobsen2023dissipative, Dlapa:2024cje}, and now results at 5PM-1SF order~\cite{Driesse:2024xad, Driesse:2024feo, Bern:2024adl}. The goal of this work is to bring precision computations in the general EFT of gravity nearer to the state-of-the-art PM computations in GR.

We consider the general effective theory of gravity, with an arbitrary number of light scalar fields in addition to the metric, and we systematically include higher derivative modifications~\footnote{In this work we will not be concerned with constraints which come from understanding GR as an effective \textit{quantum} field theory. Physical principles such as unitarity, analyticity, and causality, tend to require the existence of an infinite towers of higher-spin fields with masses set by the scale $\ell^{-1}$---challenging the notion of locality for gravitating systems on length scales below $\ell$~\cite{Camanho:2014apa, Bern:2021ppb, Caron-Huot:2022ugt, Caron-Huot:2024lbf}. While such results are certainly of fundamental importance,  we will not address them in this work, and will instead consider the EFT as just a method for parameterizing predictions of classical gravity theories}. We study the conservative scattering of two compact bodies by computing the radial action to 3PM order, i.e. two-loops. In the metric sector we retain all contributions through to sixth-power of the new physics scale $\ell$. With the scalars we include all terms through to $\ell^4$ Some of the higher-curvature operators we include have been previously studied at 2PM order~\cite{Cristofoli:2019ewu, Brandhuber:2019qpg, Emond:2019crr, AccettulliHuber:2020dal, Bern:2020uwk, Cheung:2020gbf}. Dynamics at 3PM have also been studied for spinning bodies in dynamical Chern-Simons theory~\cite{Bhattacharyya:2024kxj}.  However, we find disagreement with their spinless limit. Scalar-tensor theory, EdGB, and eight-derivative gravitational corrections have been studied to the third order in the post-Newtonian approximation~\cite{Bernard:2018hta, Bernard:2022noq, Jain:2023fvt, Jain:2023vlf, Julie:2022qux, Endlich:2017tqa},  and where our results overlap we have perfect agreement.

\vspace{6pt}
\noindent {\bf Beyond General Relativity:}
General relativity is a theory which is both generally covariant and locally Lorentz invariant. Assuming that GR is an effective theory that is a good approximation at distance scales larger than $\ell$, we would like to parameterize the most general departures from GR as one probes the scale $\ell$. In particular, we would like an effective theory for computing corrections perturbatively in powers of $\ell/r_S$, where $r_S$ is the Schwarzschild radius of a compact object. A completely general parameterization is beyond the scope of this work, however we will provide a sufficiently general parameterization to cover many models of interest.

We will limit ourselves to an effective theory which respects general covariance and locally Lorentz invariance. We allow for an arbitrary number of massless scalar fields, which we assume to interact with Standard Model fields only with gravitational strength, and we do not include vector fields.

Through $\mathcal{O}(\ell^{4})$, the most general effective gravitational theory compatible with these requirements is
\begin{align}\label{eq:effAction}
S=\frac{1}{16\pi G}\int d^{4}x \sqrt{-g}\bigg[&-R + \delta^{ab}\,\partial_\mu\phi_{a}\partial^\mu\phi_{b} \nonumber \\
&+ \ell^{2}\mathcal{L}^{\textrm{scalar}}+\ell^{4}\mathcal{L}^{\textrm{dim 6}}\bigg]\,.
\end{align}
where
\begin{align}\label{eq:scalarSector}
\mathcal{L}^{\textrm{scalar}}&=  c_{5,1}\phi_{1}\mathcal{C}+c_{5,2}(\phi_{1}\sin\chi+\phi_{2}\cos\chi)\tilde{\mathcal{C}}\,,
\end{align}
and
\begin{align}
\mathcal{L}^{\textrm{dim 6}}&=c_{6,1}\tensor{R}{_\mu_\nu^\sigma^\rho}\tensor{R}{_\sigma_\rho^\alpha^\beta}\tensor{R}{_\alpha_\beta^\mu^\nu}\nonumber+ c_{6,2}\tensor{R}{_\mu^\sigma_\nu^\rho}\tensor{R}{_\sigma^\alpha_\rho^\beta}\tensor{R}{_\alpha^\mu_\beta^\nu} \nonumber \\ 
&+c_{6,3}\tensor{\tilde{R}}{_\mu_\nu^\sigma^\rho}\tensor{R}{_\sigma_\rho^\alpha^\beta}\tensor{R}{_\alpha_\beta^\mu^\nu}\,,
\end{align}
with $\tilde{R}_{\mu\nu\sigma\rho}=\frac{1}{2}\tensor{\epsilon}{_\mu_\nu^\alpha^\beta}R_{\alpha\beta\sigma\rho}$, and  $\mathcal{C}=R_{\mu\nu\sigma\rho}R^{\mu\nu\sigma\rho}$, and $\tilde{\mathcal{C}}=R_{\mu\nu\sigma\rho}\tilde{R}^{\mu\nu\sigma\rho}$. For the metric sector, we will also include the most general action through $\mathcal{O}(\ell^{6})$ \cite{Endlich:2017tqa},
\begin{align}
\ell^{6}\mathcal{L}^{\textrm{dim 8}}=c_{8,1}\,\mathcal{C}^{2}+c_{8,2}\,\tilde{\mathcal{C}}^{2}+c_{8,3}\,\mathcal{C}\tilde{\mathcal{C}}\,,
\end{align}
 There are also scalar interactions to account for at this order. However we will will limit our analysis of scalars to $\mathcal{O}(\ell^{4})$. The $c_{i,j}$, as well as $\chi$, are dimensionless constants determined by the UV gravity theory. We've omitted terms which are total derivatives in four dimensions, and eliminated most terms which are redundant under field redefinitions.

In four dimensions the $c_{6,2}$ term is redundant with the $c_{6,1}$ term~\cite{vanNieuwenhuizen:1976vb}. The vanishing of the cubic Lovelock density implies that it can be written as a linear combination of the $c_{6,1}$ term and terms which involve the Ricci curvature and can therefore be eliminated by field redefinition~\cite{Cano:2019ore}. This redundancy can be used to set $c_{6,2}=0$. However, we will keep its value arbitrary to allow for a consistency check on later results. There is an analogous parity-odd term which has already been set to zero.
At the eight-derivative order we've only written linearly independent terms~\cite{Endlich:2017tqa}.

In the scalar sector, we have used field redefinitions to orthonormalize the kinetic term. The remaining $SO(N)$ symmetry allows one to rotate the field basis such that only one field, $\phi_{1}$, couples to $\mathcal{C}$. The remaining $SO(N-1)$ rotations allow one to rotate further so that only $\phi_{1}$ and $\phi_{2}$ couple to $\tilde{\mathcal{C}}$ \cite{Cano:2019ore}.

Since we've effectively set the Ricci tensor to zero in higher-order terms by field redefinition, the Kretschmann scalar $\mathcal{C}$ is equivalent to the Euler density, and $c_{5,1}$ corresponds to the coupling in Einstein-dilaton-Gauss-Bonnet  (EdGB) gravity~\footnote{To order $\ell^4$, there is no distinction between dilaton-Gauss-Bonnet, truncated dilaton-Gauss-Bonnet, and a general Einstein-scalar-Gauss-Bonnet theory.}. The angle $\chi$ is a free parameter characterizing parity breaking in the scalar sector. The pseudoscalar $\phi_{2}$ can be identified as the axion in dynamical Chern-Simons (dCS) gravity~\cite{Jackiw:2003pm, Alexander:2004us}.

The scalar charge induced by the EdGB and dCS couplings on gravitational solutions will be of order $\mathcal{O}(\ell^{2})$, so we can power count both the dimension-5 operators as well as the dilaton and axion kinetic terms as $\mathcal{O}(\ell^{4})$. To reiterate, if $\ell=0$, the compact objects we're interested in would not support any scalar hair i.e. source the scalar field. A nontrivial source for the scalar field starts only at $\mathcal{O}(\ell^{2})$. We'll comment on the scaling of the remaining $N-2$ scalars momentarily. Scalar-scalar interactions such as $\phi^{3}$, $(\partial\phi)^{2}$, etc. are of the correct derivative order, but since $\phi\sim\ell^{2}$, they power count too high in $\ell$ to be relevant. Hence, we can truncate the action to quadratic order in $\phi_{a}$. Note, this effective Lagrangian then includes multi-scalar-tensor theories, with generic potential $V(\phi_{a})$ and Brans-Dicke function $\omega(\phi_{a})$, provided we understand that the dynamical fields here are perturbatively expanded about some constant background values $\bar{\phi}_{a}.$ The only nongeneric assumption we've made is that there are no mass terms~\footnote{If relevant, such terms would render the forces mediated by these scalars to be short ranged and would exponentially suppress the modifications to gravity.}.

We have not yet discussed the compact objects whose gravitational dynamics we will study. No-hair theorems prevent black holes from carrying charge under the scalar fields $\phi_{n}$ for $n>2$~\cite{Hawking:1972qk}, while compact matter can carry scalar charge in scalar-tensor theories \cite{Eardley:1975, PhysRevLett.70.2220}. Conversely, compact objects other than black holes (eg. neutron stars) do not have charge induced by EdGB and dCS couplings ~\cite{Yagi:2015oca}, whereas black holes do~\cite{PhysRevD.90.124063}.  With the normalization of our fields in \cref{eq:scalarSector}, the asymptotic form of the scalar field outside an object of mass $m$ is
\begin{align}
\phi = d\frac{Gm}{r}+\mathcal{O}(r^{-2})\,.
\end{align}
For a black hole in EdGB theory, $d=c_{5,1}\ell^2/(Gm)^2$, and $c_{5,1}\ell^2$ is bounded to be less than $(0.43\,G M_\odot)^2$, so it is safely small compared to the size of astrophysical black holes and neutron stars~\cite{Sanger:2024axs}.  For an object in scalar-tensor theory, in standard Brans-Dicke scalar-tensor parameters,
\begin{equation}
d=\frac{(2s-1)}{4}\left(\frac{2}{3+2\omega_{0}}\right)^{1/2}\,,
\end{equation}
where $s$ is the leading ``sensitivity'' which is $0.5$ for black holes and expected to be in the range $\mathcal{O}(0.1-0.5)$ for neutron stars~\cite{Hawking:1972qk, Eardley:1975}. The Brans-Dicke coupling constant is observationally bounded $\omega_{0}>10^{4}$~\cite{DeFelice:2010aj}, and so the scalar charge $d$ is necessarily small in dimensionless units. To parallel the EdGB discussion, we can define a ``new physics'' length scale for scalar-tensor theory,
\begin{align}
    \ell_{ST}^{2}=\left(\frac{1}{3+2\omega_{0}}\right)^{1/2} (Gm)^2\,,
\end{align}
where $m$ is the mass of the lightest observed strong gravity binary system, and the power counting argument for our effective action will continue to hold.

Given this effective theory of gravity, valid for scales longer than $\ell$, we would now like to discuss the effective field theory for compact objects in such a theory. That is, the worldline theory valid for length scales $b$ much larger than the size of the compact object body, i.e. $b \gg Gm \gg \ell$. In gravity this was worked in detail in \cite{NRGR} (see \cite{Porto:2016pyg, Levi:2018nxp, goldberger2023effective} for review). An effective description of compact objects in scalar-tensor gravity has been long known~\cite{Eardley:1975}, although a systematic understanding of the point particle approximation, e.g. the divergences it introduces and the renormalization it requires, came only after \cite{NRGR}. To the order we're working, we need only the leading sensitivity i.e. the linear coupling of the compact object to the scalars. Scalar ``tidal'' interactions described by $\phi^2$ etc. on the worldline scale as too high a power in $\ell$. 

The worldline action for two compact bodies is \footnote{Note: Our definition of the tidal couplings differs from convention by a factor of the mass.}
\begin{align}
S&=\sum_{n=1,2}m_{n}\int d\lambda \sqrt{g_{\mu\nu}\dot{x}^{\mu}_n\dot{x}^{\nu}_n}\bigg[-1+d^{(n)}_a\phi_{a} \nonumber \\
&+c^{(n)}_{E}E_{\mu\nu}E^{\mu\nu}+c^{(n)}_{B}B_{\mu\nu}B^{\mu\nu}\bigg]\,.
\end{align}
Here we've also include the leading finite size (tidal) couplings to the gravitational field---$E_{\mu\nu}$ and $B_{\mu\nu}$ being the electric and magnetic parts of the Weyl tensor. In vacuum the field redefinition used to eliminate the higher-derivative $c_{6,2}$ is innocuous, however in the presence of compact objects it mixes with the tidal couplings~\cite{Bern:2020uwk, AccettulliHuber:2020dal}. Tidal effects have been well studied, including at the two-loop order we study in this work~\cite{Kalin:2020lmz}. We include tidal contributions only for a consistency check on the novel calculation, i.e. that a redefinition of $c_{6,1}, c_{E}, c_{B}$ can eliminate $c_{6,2}$ from physical observables. 

In the effective action \cref{eq:effAction} we have a number of parity-breaking terms, parameterized by the couplings $\sin\chi,c_{6,3},c_{8,3}$. Furthermore, if $c_{5,2}$ is nonzero, and we treat $\phi_{2}$ as a pseudoscalar field, then the worldline coupling $d_{2}\phi_{2}$ is also parity breaking. Our results for the two-loop scattering of spinless bodies will be completely insensitive to parity breaking couplings. This can be anticipated before computing. We can treat the couplings $\sin\chi,c_{6,3},c_{8,3},d_{2}$ as spurions for parity breaking, ie. assign them odd parity transformations to restore symmetry to the theory. Since we consider the scattering of spinless bodies, the observables will then be invariant under parity. This is impossible at linear order in $\sin\chi,c_{6,3},c_{8,3}$, so such linear contributions must vanish. At quadratic order in couplings we can have $\sin^{2}\chi$ and  $d_{2}^{2}$ terms, however the $\sin^{2}\chi$ is necessarily accompanied by $\cos^{2}\chi$ due to $\phi_{1}$ exchange. The angle $\chi$ then completely drops out of the observables at this order, and $d_{2}\phi_{2}$ contributes only in a parity-even manner. There is also, in principle, a parity breaking tidal-term $E_{\mu\nu}B^{\mu\nu}$ on the worldline at this order, which we have omitted for the same reasons. To be sensitive to parity breaking terms, one needs to proceed to a higher order in these couplings or include spin on the worldline. We leave this for future work.

\noindent {\bf Radial Action:} We computed perturbative scattering of two compact objects in the above effective theory to third post-Minkowskian order. While some of the couplings, particularly the scalar charges and tidal coefficients, implicitly depend on $G$ after the appropriate matching calculation has been performed, here we are computing to an \emph{explicit} order $G^{3}$, i.e. two-loop order. Similarly, when we refer below to power counting in the masses of the objects,  we are referring to explicit orders in the mass parameter and are not including the intrinsic dependence of couplings on mass implied by matching. Since we ultimately compute the full PM result, this is just a choice of bookkeeping. 

We consider objects with asymptotic velocities $u_{1,2}^{\mu}$ with relative boost $\gamma=u_{1}\cdot u_{2}$ and woth impact parameter $b$,   and compute the radial action  $i_{r}(\gamma,b)$. The radial action is a generating function from which all gauge invariant observables for the conservative two-body problem follow. The scattering parameters are expressible in terms of binding energy $\mathcal{E}$ and angular momentum $J$, via
\begin{align}\label{eq:orbitalParameters}
J&= Mb\frac{\nu(\gamma^2-1)}{1+2\nu(\gamma-1)} \, , \nonumber \\
E&=M(1+\nu\mathcal{E})\,,\nonumber \\
\gamma&=1+\mathcal{E}+\frac{\nu}{2}\mathcal{E}^{2}\,,
\end{align}
where $M$ and $\nu$ are the total mass and symmetric mass ratio. Observables for the bound system follow immediately after analytically continuing $i_{r}(J,\mathcal{E})$ to negative $\mathcal{E}$~\cite{Kalin:2019rwq, Kalin:2019inp}. In the interest of space, we will present only the radial action.

The radial action in this setup is the value of the action functional evaluated on the solution to the equations of motion, i.e. Hamilton's principal function with energy and angular momentum, rather than coordinates, being specified asymptotically. In field theory parlance, it is the on-shell action. On-shell actions can be computed perturbatively by iteratively solving the equations of motion and inserting the solution into the action, or equivalently, by summing a set of Feynman diagrams with prescribed momentum transfer and performing a Fourier transform from momentum transfer back to impact parameter space (see, e.g. \cite{Porto:2016pyg, Kalin:2020mvi} for details). 

The leading 1PM result for the EFT considered herein, is given the sum of a tree-level exchange of a graviton and a scalar. Omitting the IR divergence in the Coulomb logarithm, this leading result is
\begin{align}
i_{r}=-Gm_{1}m_{2}\log(b)\frac{(4\gamma^2-2)+\vec{d}^{\,(1)}\cdot \vec{d}^{\,(2)}}{(\gamma^{2}-1)^{1/2}}\,.
\end{align}

To proceed beyond 1PM we used the Effective Field Theory for Extreme Mass Ratio Binaries \cite{Cheung:2023lnj, Cheung:2024byb} as an organizational tool.  However, our computation captures the full 3PM results. Through 4PM the radial action is determined solely by two contributions: the probe limit in which one body is orbiting in the fixed background of the other and the leading correction to this limit referred to as the 1SF (``self-force'') contribution. When necessary, one must symmetrize the lower SF order radial action appropriately over the labels $1\leftrightarrow 2$ to obtain the full result.  The full details of the 3PM 1SF computation in GR and electrodynamics are given in~\cite{Cheung:2024byb, Wilson-Gerow:2023syq}.

The probe limit is straightforwardly computed by evaluating a radial action integral (see \cite{Damour:1988mr, Kalin:2019rwq}). We computed spherically symmetric solutions in the modified theory perturbatively in $\ell$, solved for the radial momentum of a probe particle in terms of the conserved energy and angular momentum, and then integrated over the scattering orbit (see Appendix 2). The 1SF contributions require the evaluation of Feynman loop diagrams. We treat one of the bodies, 1, as the fixed background for the SF expansion and then proceed as outlined in~\cite{Cheung:2023lnj, Cheung:2024byb}.

\begin{figure}
\begin{centering}
\includegraphics[width=0.33\textwidth]{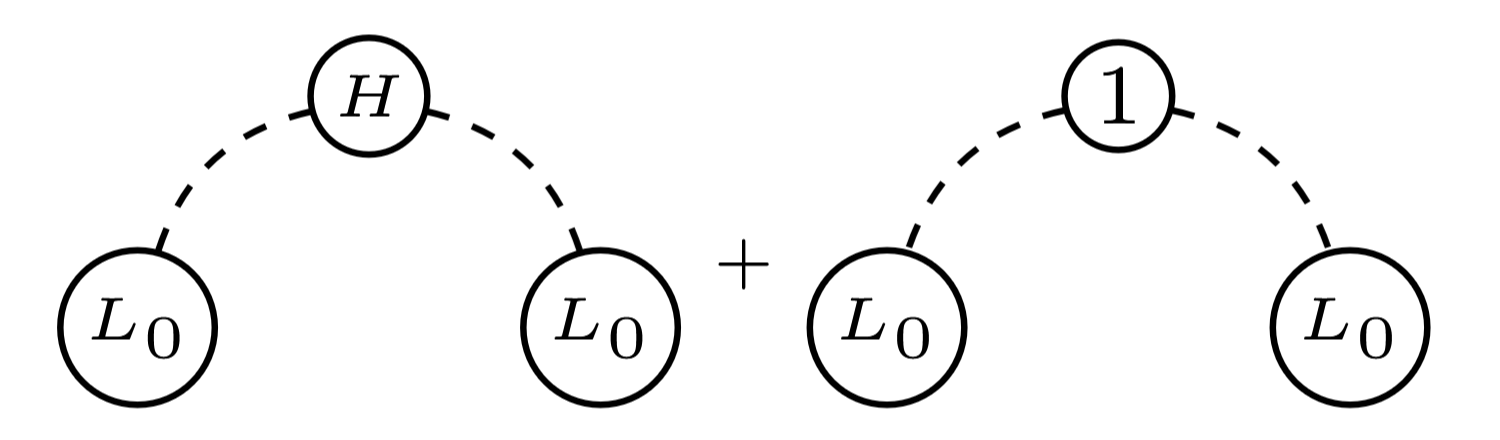}
\caption{Diagram structures for the 1SF-2PM radial action. Vertices describe light-particle probe-motion sources, heavy particle recoil/tidal operators, or linearized background field insertions, notated by $L_{0}, H, 1$ respectively. Dashed lines can be gravitons or scalars.}\label{fig:2PMdiagrams}
\end{centering}
\end{figure}

\begin{figure}
\begin{centering}
\includegraphics[width=0.35\textwidth]{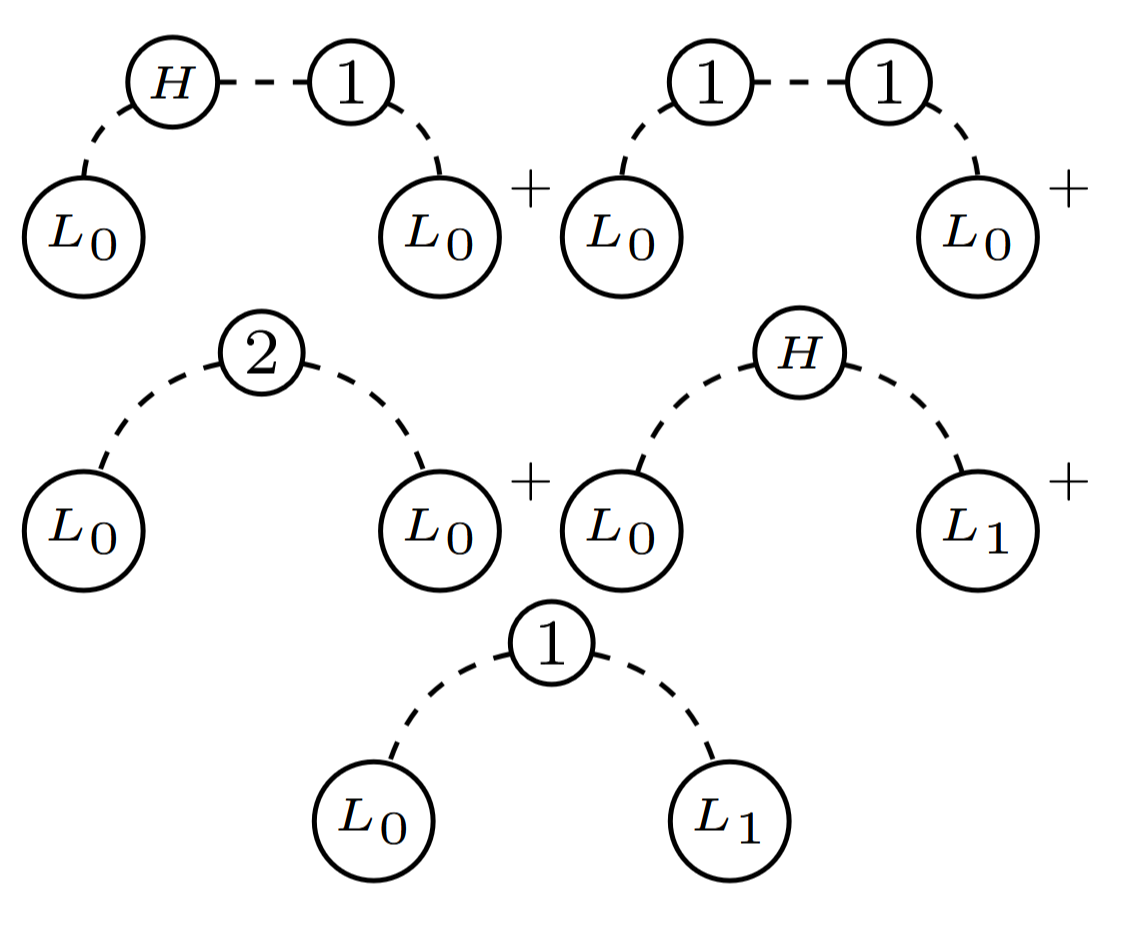}
\caption{Diagram structures for the 1SF-3PM radial action. The background field vertex, 2, is a one-loop sub-diagram insertion. $L_{1}$ is a tree-level corrected probe-motion source.}
\label{fig:3PMdiagrams}
\end{centering}
\end{figure}

The relevant diagrams are the same form as in GR (see ~\cref{fig:2PMdiagrams}), however we must sum over the identities of all of the particles running through the graph. Since spinless bodies do not source the axion field, the dCS vertex does not contribute at this order.  Each of these diagrams has been computed and we present the 2PM result as
\begin{align}\label{eq:2PMir}
i_{r}=\frac{G^2m^2_{1}m_{2}\pi}{b}\frac{h_{1}}{(\gamma^{2}-1)^{1/2}} +(1 \leftrightarrow 2)\,
\end{align}
where for each theory $h_{1}$ is the product of the ``new-physics'' couplings with a polynomial in the relative boost factor $\gamma$ (see \cref{table:2PMresuls} in Appendix 1 for the explicit form).

At 2PM order, we computed the probe and 1SF contributions separately, as described above, without symmetrizing over the labels. It is a nontrivial check that these two contributions to the radial action have identical forms, i.e. that the result \cref{eq:2PMir} is indeed symmetric under interchanging the labels 1 and 2. Since 2PM conservative dynamics are determined by probe physics and do not necessitate Feynman integrals the main result of this work is the following 3PM contribution. 

The radial action at 3PM can be written as
\begin{align}\label{eq:3PMir}
i_{r}=&\frac{G^3m^3_{1}m_{2}}{b^2}\frac{h_{2}}{(\gamma^{2}-1)^{5/2}} +(1 \leftrightarrow 2) \nonumber\\
+&\frac{G^3m^2_{1}m^2_{2}}{b^2}\left(\frac{h_{3}}{(\gamma^{2}-1)^{5/2}}+\frac{h_{4}\arccosh{\gamma}}{(\gamma^{2}-1)^{3}}\right)\,,
\end{align}
where the $h_2$ and $h_{3},h_{4}$ polynomials (see App. 1) describe contributions from the probe limit and 1SF respectively, and the symmetrization is intended only for the probe terms.  At this order we opted to only directly compute the probe and 1SF contributions, and to infer 2SF from the probe result. The 1SF Feynman diagrams (\cref{fig:3PMdiagrams}) were assembled using xAct, and were integral-reduced using LiteRed and FIRE codes~\cite{Lee:2013mka, Smirnov:2019qkx}. The integration details can be found in ~\cite{Parra-Martinez:2020dzs, Dlapa:2023hsl}.

 The result passes a variety of consistency checks.  We include the Einstein gravity tidal contribution, and it agrees with known results~\cite{Kalin:2020lmz}.  Ignoring higher-derivative couplings, and setting the scalar charge of one of the bodies to zero, agrees with the scalar 1SF results in~\cite{Cheung:2023lnj, Cheung:2024byb, Barack:2023oqp}. In the computations we used a general $R_{\xi}$ gauge-breaking term, $\mathcal{L}_{\textrm{g.b.}}=\xi(32\pi G)^{-1}\sqrt{-g}F_{\mu}F^{\mu}$, with $F_{\mu}=\bar{\nabla}_{\nu}\delta g_{\mu}^{\nu}-\frac{1}{2}\bar{\nabla}_{\mu}\delta g$, and the result is independent of $\xi$. Furthermore, although we did not write it explicitly, we also included a set of redundant operators including $\phi_{1}R^{2}, \phi_{1}R_{\mu\nu}R^{\mu\nu}$, and verified that the result is independent of these interactions. Taking the static limit, the effective potential implied by the quadratic-curvature terms $h_{1}^{\textrm{dim 8}}$ agrees precisely with  \cite{Endlich:2017tqa}. Additionally, for each $j$ we can take the $h_j^{\textrm{dim 6}}+h_j^{\textrm{tidal}}$ and verify that a redefinition of couplings
\begin{equation}
c_{6,1}\rightarrow c_{6,1}-\frac{1}{2}c_{6,2}\qquad c_{E/B}^{(n)}\rightarrow c_{E/B}^{(n)}\pm\frac{3}{2}c_{6,2}\ell^{4}\,
\end{equation}
will eliminate $c_{6,2}$ from the result, confirming that it was indeed describing a redundant operator. 

Finally we were able to perform nontrivial checks on our scalar tensor theory results. By expanding in a nonrelativistic limit and computing the scattering angle from an effective-one-body Hamiltonian~\cite{Julie:2017pkb, Julie:2022qux, Jain:2022nxs, Jain:2023vlf}, we were able to compare with known 3PN results. The 3PM calculation does not capture $\mathcal{O}(G^4)$ contributions,  where tail effects arise in scalar tensor theory.  However where there is overlapping validity we found agreement~\footnote{When comparing it was crucial to truncate their calculation to quadratic order in scalar charge to be consistent with the approximation we've taken here.}.

Our dCS results disagree with \cite{Bhattacharyya:2024kxj} who computed the two-loop eikonal phase in dCS theory for spinning bodies, from which the scattering angle is readily derived. In the spinless limit black holes do not source the dCS scalar field since spherically symmetric GR solutions are not modified in dCS theory. Consequentially the ``sensitivity'', which describes the linear dependence of the compact object's mass on the value of the scalar field, should be zero. Taking this limit in in the result of \cite{Bhattacharyya:2024kxj}, along with the vanishing of the spin, their two-loop eikonal phase vanishes. However, in this work we find a nonzero result.

In dynamical settings such as the scattering problems there is \emph{orbital} angular momentum which breaks parity and allows for coupling to the dCS axion. This is seen at two-loops where there is axion exchange along the internal line of the diagram with two linearized background insertions, even when the bodies are spinless. That is, a diagram with H-topology where only the horizontal line is an axion and the rest are gravitons. Such a diagram leads to the nonzero result we found in this work, and appears to have been omitted by \cite{Bhattacharyya:2024kxj}.

\vspace{6pt}
\noindent {\bf Conclusions}: In this work we studied the binary dynamics of compact objects in a general modification to Einstein gravity which includes a metric field, an arbitrary number of massless scalars, and includes higher derivative interactions between them. We computed the radial action in a post-Minkowskian expansion to $O(G^{3}\ell^{6})$, in the purely metric theory, and to $O(G^{3}\ell^{4})$ in the theory with scalars. In addition to higher-curvature gravity, this general effective theory includes, as special cases, multi-tensor-scalar, Einstein-dilaton-Gauss-Bonnet, and dynamical Chern-Simons theories. This work goes beyond the 2PM work previously done in binary dynamics in effective theories, and appends to the potentials in previous 2PN work, all $(v/c)^2$ contributions.

A primary motivation of this work was to make progress toward the goal of supplementing the recent successes in PM computations in GR with commensurate results for a general deviation from Einstein gravity that is constrained by known physical principles. Incorporating such results into waveform generation will provide a physically motivated parameter space against which GR can be tested as a null hypothesis when comparing to observational data.  We've been sufficiently general in this work that we can make the rather strong claim: if GW data were to be inconsistent with any choice of parameters in~\eqref{eq:effAction}, then gravity would not be described by a generally covariant theory of a locally Lorentzian metric along with any number of massless scalar modes. 

There are a number of immediate directions for future work. Increasing the PM order is an obvious route, to follow the state-of-the-art progress in GR. Many of the results obtained here scale as $b^{-6}$, though, which would appear as a 6PM order modification in GR. Another important route would be to include spin.  This is important to assess whether signatures of certain GR modifications are degenerate with spin-contributions. Moreover, while dCS interactions contributed only at two-loop for spinless bodies, for spinning bodies dCS interactions are enhanced to tree-level. Furthermore, parity breaking higher-curvature operators contribute once the bodies have spin. Superradiant clouds around spinning black holes offer yet another exciting  motivation.

The work here was focused on the conservative sector. Scalar charged objects experience enhanced radiative losses---in the post-Newtonian limit this fact already places tight bounds on new physics~\cite{will:1977, will:1989, PhysRevD.50.6058, PhysRevD.54.1474, Damour:1998jk}---and it would be important to properly characterize this dissipation in PM computations. 

Finally, it would be useful to import the theoretical data obtained here into an effective one-body model, or another analogous resummation tool, so that it can be readily used to place constraints on departures from Einstein gravity once observational data sensitive to the early phases of eccentric binary inspirals becomes available. Unfortunately, though, in \cref{table:h2,table:h34} we can see that none of the beyond-GR contributions grow with $\gamma$ faster than the GR contributions. This suggests that neither highly eccentric binaries nor unbound scattering encounters will offer enhanced observational sensitivity to such corrections.

\vspace{6pt}

\noindent {\bf Acknowledgments:} This work is supported
by the US Department of Energy (HEP) Award DE-SC0013528, as well as a President's Postdoctoral Fellowship at Carnegie Mellon University. Some of this work was completed at Caltech, while supported by the Department of Energy (Grant No.~DE-SC0011632), the Walter Burke Institute for Theoretical Physics, a Presidential Postdoctoral Fellowship, and the Simons Foundation (Award Number 568762).

\appendix

\newpage

\section{Appendix 1: Velocity polynomials}

The 2PM velocity polynomials are
\begin{table}[H]
\setlength{\tabcolsep}{5pt} 
\renewcommand{\arraystretch}{2}
\begin{tabular}{|c|}
\hline
\parbox{0.8\columnwidth}{
\begin{align}
    h_1^{\textrm{GR}} &= \frac{3}{4} \left(5 \gamma ^2-1\right) \nonumber\\
    h_1^{\textrm{tidal}} &= \frac{c^{(1)}_E}{b^{4}}\frac{9}{64} \left(35 \gamma ^4-30 \gamma ^2+11\right)\nonumber\\
    &+\frac{c^{(1)}_B}{b^{4}}\frac{9}{64}\left(35 \gamma ^4-30 \gamma ^2-5\right)\nonumber\\
    h_1^{\textrm{ST}}&=\vec{d}^{\,(1)}\cdot \vec{d}^{\,(2)}-|\vec{d}^{\,(1)}|^2\frac{1}{8} \left(\gamma ^2-1\right)  \nonumber\\
    h_1^{\textrm{EdGB}}&=-\frac{c_{5,1}\ell^2}{2b^2} \left(d^{\,(2)}_{1}+(3\gamma ^2-1)d_{1}^{\,(1)}\right) \nonumber\\
    h_1^{\textrm{dCS}}&=0 \nonumber\\
    h_1^{\textrm{dim 6}}&=-\frac{c_{6,1}\ell^4}{b^{4}} \frac{27}{4} \left(\gamma ^2-1\right)-\frac{c_{6,2}\ell^4}{b^{4}}\frac{27 \gamma ^2 }{8 } \nonumber \\ 
\end{align}}  \\
\hline
\end{tabular}
\caption{2PM velocity polynomials}
\label{table:2PMresuls}
\end{table}
The 3PM-0SF velocity polynomials are
\begin{table}[H]
\setlength{\tabcolsep}{5pt} 
\renewcommand{\arraystretch}{2}
\begin{tabular}{|c|}
\hline
\parbox{0.8\columnwidth}{
\begin{align}
    h_2^{\textrm{GR}} =& \frac{1}{3} \left(64 \gamma ^6-120 \gamma ^4+60 \gamma ^2-5\right) \nonumber\\
    h_2^{\textrm{tidal}} =&\frac{c_{E}^{(2)}}{b^{4}}\frac{32}{35}\left(\gamma^2-1\right) \left(8 \gamma ^2 \left(20 \gamma ^4-24 \gamma ^2+9\right)-5\right) \nonumber\\
    +&\frac{c_{B}^{(2)}}{b^{4}}\frac{64}{35} \left(\gamma ^2-1\right)^2\left(80 \gamma ^4-16 \gamma ^2-1\right)  \nonumber\\
    h_2^{\textrm{ST}}=& \frac{1}{6} \left(44 \gamma ^4-70 \gamma ^2+23\right) \vec{d}^{\,(1)}\cdot\vec{d}^{\,(2)} \nonumber \\
    -&\frac{2}{3} \left(\gamma ^2-1\right)^2 \left(2 \gamma^2-1\right) |\vec{d}^{\,(1)}|^2   \nonumber\\
    h_2^{\textrm{EdGB}}=&-\frac{c_{5,1}\ell^2}{b^2}\frac{4}{15} \left(\gamma ^2-1\right) \left(114 \gamma ^4-97 \gamma ^2+13\right) d^{(1)}_1\nonumber\\
    &-\frac{c_{5,1}\ell^2}{b^2}\frac{4}{15} \left(\gamma ^2-1\right)\left(28 \gamma ^2-13\right)   d^{(2)}_1 \nonumber \\
    h_2^{\textrm{dCS}}=&0 \nonumber\\
    h_2^{\textrm{dim 6}}=&- \frac{c_{6,1}\ell^4}{b^4}\frac{2048}{105} \left(\gamma ^2-1\right)^2  \left(8 \gamma ^2-1\right)\nonumber \\
    &- \frac{c_{6,2}\ell^4}{b^4}\frac{16}{105} \left(\gamma ^2-1\right)\left(512 \gamma ^4-198 \gamma
   ^2+1\right) \nonumber \\ 
   h_2^{\textrm{dim 8}}=&-\frac{c_{8,1}\ell^6}{b^{6}}\frac{8192}{35}(\gamma^2-1)^2(3\gamma^2-1)\,
\end{align}}  \\
\hline
\end{tabular}
\caption{3PM-0SF  velocity polynomials}
\label{table:h2}
\end{table}

The 3PM-1SF velocity polynomials are
\begin{table}[H]
\setlength{\tabcolsep}{5pt} 
\renewcommand{\arraystretch}{2}
\begin{tabular}{|c|}
\hline
\parbox{0.8\columnwidth}{
\begin{align}
    h_3^{\textrm{GR}} =&\frac{2}{3} \gamma  \left(36 \gamma ^6-114 \gamma ^4+132 \gamma ^2-55\right)  \nonumber\\
    h_4^{\textrm{GR}} =& -4 \left(\gamma ^2-1\right)^2 \left(4 \gamma ^4-12 \gamma ^2-3\right) \nonumber\\
    h_3^{\textrm{tidal}} =&\frac{c_{E}^{(1+2)}}{b^{4}}\frac{16}{5} \gamma  \left(32
   \gamma ^8-104 \gamma ^6+804 \gamma ^4+9426 \gamma ^2+4047\right)  \nonumber\\
    +&\frac{c_{B}^{(1+2)}}{b^{4}}\frac{16}{5} \gamma  \left(32 \gamma ^8-104 \gamma ^6+784 \gamma ^4+9466 \gamma ^2+4012\right)   \nonumber\\
    h_4^{\textrm{tidal}} =&-\frac{c_{E}^{(1+2)}}{b^{4}}48 \left(440 \gamma ^4+474 \gamma ^2+33\right) \nonumber\\
    &-\frac{c_{B}^{(1+2)}}{b^{4}}48 \left(440 \gamma ^4+474 \gamma ^2+32\right) \nonumber\\
    h_3^{\textrm{ST}}=&-(|\vec{d}^{\,(1)}|^2+|\vec{d}^{\,(2)}|^2) \frac{1}{3} \gamma  \left(2 \gamma ^6-3 \gamma ^4+1\right) \nonumber \\
    &+\vec{d}^{\,(1)}\cdot \vec{d}^{\,(2)}\gamma  \left(10 \gamma ^4-14 \gamma ^2+3\right)   \nonumber \\
    h_4^{\textrm{ST}}=&-\vec{d}^{\,(1)}\cdot \vec{d}^{\,(2)}8 \left(\gamma ^2-1\right)^3   \nonumber \\
    h_3^{\textrm{EdGB}}=&\frac{c_{5,1}\ell^2}{b^{2}}(d^{(1)}_1+d^{(2)}_1) \frac{4}{3} \gamma  \left(\gamma ^2-1\right) \left(16 \gamma ^4-22 \gamma ^2-243\right)\nonumber\\
    -&\frac{c_{5,1}^2\ell^4}{b^{4}}\frac{128}{3} \gamma  \left(44 \gamma ^6-94 \gamma ^4+356 \gamma ^2+351\right) \nonumber \\
    h_4^{\textrm{EdGB}}=&-\frac{c_{5,1}\ell^2}{b^{2}}(d^{(1)}_1+d^{(2)}_1)16 \left(4 \gamma ^6-23 \gamma ^4+14 \gamma ^2+5\right) \nonumber\\
    &+\frac{c_{5,1}^2\ell^4}{b^{4}}128 \left(8 \gamma ^8-8 \gamma ^6+36 \gamma ^4+166 \gamma ^2+17\right) \nonumber \\
    h_3^{\textrm{dCS}}=&-\frac{c_{5,1}^{2}\ell^4}{b^{4}}\frac{16}{3} \gamma  \left(22 \gamma ^6-47 \gamma ^4+124 \gamma ^2+126\right) \nonumber\\
    h_4^{\textrm{dCS}}=&\frac{c_{5,2}^{2}\ell^4}{b^{4}}16 \left(4 \gamma ^8-4 \gamma ^6+9 \gamma ^4+60 \gamma ^2+6\right) \nonumber\\
    h_3^{\textrm{dim 6}}=&\frac{c_{6,1}\ell^4}{b^{4}}256 \gamma  \left(2 \gamma ^6-7 \gamma ^4+2 \gamma ^2-42\right) \nonumber\\
    +&\frac{c_{6,2}\ell^4}{b^{4}}16 \gamma  \left(16 \gamma ^6-68 \gamma ^4+40 \gamma ^2-357\right)\nonumber\\ 
    h_4^{\textrm{dim 6}}=&\frac{c_{6,1}\ell^4}{b^{4}}2304 \left(4 \gamma ^2+1\right)+\frac{c_{6,2}\ell^4}{b^{4}}144 \left(32 \gamma ^2+9\right) \nonumber\\
     h_3^{\textrm{dim 8}}=& h_4^{\textrm{dim 8}} = 0
\end{align}}  \\
\hline
\end{tabular}
\caption{3PM-1SF velocity polynomials}
\label{table:h34}
\end{table}
where we've defined
\begin{equation}
c_{E}^{(1+2)}=c_{E}^{(1)}+c_{E}^{(2)}\,,
\end{equation}
for both the electric and magnetic terms.

\section{Appendix 2: Probe radial action}

To compute to 0SF results, as well as the background field insertions in the 1SF diagrams, we need to first find solutions to the equations of motion in the gravitational effective theory. We treat particle 1 as the background, and particle 2 as the probe. To proceed we assume a spherically symmetric solution
\begin{align}
ds^2&=A(r)dt^2-\frac{dr^2}{B(r)}-r^2(d\theta^2+\sin^2\theta d\phi^2)\,,\\
\phi_a&=d^{(1)}_a C_a(r)\,,
\end{align}
with the deviations from Schwarzschild described by a perturbative series
\begin{align}
A(r)&=1-\frac{r_{S}}{r}+\sum_{n=2}a_{n}\left(\frac{r_{S}}{r}\right)^n\,,\nonumber\\
B(r)&=1-\frac{r_{S}}{r}+\sum_{n=2}b_{n}\left(\frac{r_{S}}{r}\right)^n\,,\nonumber \\
C(r)&=\sum_{n=1}C_{n}\left(\frac{r_S}{r}\right)^n\,,
\end{align}
where $r_S=2Gm_{1}$.  We then solved the equations of motion in the effective theory. We solved for $C(r)$ perturbatively in powers of $r_{s}/r$, and solved for $A(r),B(r)$ perturbatively in powers of $Gm/r$ and to quadratic order in the scalar charge. When accounting for quartic curvature corrections we computed to leading order, $\ell^6$.

We can start the $a_{n},b_{n}$ series from $n=2$, without loss of generality, as this amounts to expressing all quantities in terms of the renormalized mass.

The cubic and quartic curvature corrections, to order $G^3$, give the non-zero couplings
\begin{align}
a_{6}&=\frac{9c_{6,2}}{2}\left(\frac{\ell}{r_{S}}\right)^4\,,\qquad a_{7}=(5c_{6,1}-\frac{17}{4}\,,c_{6,2})\left(\frac{\ell}{r_{S}}\right)^4\,, \nonumber \\
b_{6}&=54c_{6,1}\left(\frac{\ell}{r_{S}}\right)^4\,,\qquad b_{7}=(-49c_{6,1}+\frac{1}{4}\,,c_{6,2})\left(\frac{\ell}{r_{S}}\right)^4\,, \nonumber \\
a_{9}&=128c_{8,1}\left(\frac{\ell}{r_{S}}\right)^6\,, \qquad b_{9}=576c_{8,1}\left(\frac{\ell}{r_{S}}\right)^6\,.
\end{align}
For a general scalar-tensor theory we have the scalar profile 
\begin{align}
C_{n}&=\frac{1}{2n}\,,
\end{align}
and once including the EdGB coupling, the series for the dilaton $\phi_1$ truncates, as its coefficients read
\begin{align}
C_{n>3}=\frac{1}{2n}\left(1-\frac{4c_{6,1}\ell^{2}}{d^{(1)}_{1}r_S^2}\right)\,,
\end{align}
which vanishes upon imposing the matching condition for the charge of an EdGB black hole. The metric components to order $G^3$ and quadratic order in scalar charge is given by
\begin{align}
a_{2}&=0,\,\, a_{3}=\frac{|\vec{d}^{(1)}|^2}{48},\,\, a_{4}=\frac{|\vec{d}^{(1)}|^2}{48}+\frac{d^{(1)}_{1}c_{6,1}\ell^2}{r_S^2}\,, \nonumber \\
a_{5}&=\frac{9|\vec{d}^{(1)}|^2}{160}+\frac{d^{(1)}_{1}c_{6,1}\ell^{2}}{20r_S^2}\,, \nonumber \\
b_{2}&=\frac{|\vec{d}^{(1)}|^2}{8},\,\,b_{3}=\frac{|\vec{d}^{(1)}|^2}{16},\,\,b_{4}=\frac{|\vec{d}^{(1)}|^2}{24}+\frac{2d^{(1)}c_{6,1}\ell^2}{r_S^2}\,,\nonumber \\
b_{5}&=-\frac{|\vec{d}^{(1)}|^2}{32}+\frac{d^{(1)}_{1}c_{6,1}\ell^{2}}{20r_S^2}\,.
\end{align}

The radial momentum can be readily solved for from the on-shell condition. For a particle with scalar charge $d^{(2)}$, to leading order in charge the on-shell condition corresponds to propagation in conformally scaled metric
\begin{align}
    (1+d^{(2)}_a\phi_{a})g^{\mu\nu}p_{\mu}p_{\nu}=m^2_2\,.
\end{align}
The radial momentum is then
\begin{equation}
p_{r}^2(r)=\frac{m_{2}^{2}}{B(r)}\left(\frac{\gamma^{2}}{A(r)}-1-\frac{(\gamma^2-1)b^{2}}{r^{2}(1-d^{(2)}\cdot d^{(1)}C(r))}\right)\,,
\end{equation}
where $\gamma$ and $b$ are related to energy and angular momentum, \cref{eq:orbitalParameters}.

For a tidally coupled particle the on-shell condition reads
\begin{align}
g^{\mu\nu}p_{\mu}p_{\nu}=m^2_2\left(1-2c_{E}E_{\mu\nu}E^{\mu\nu}-2c_{B}B_{\mu\nu}B^{\mu\nu}\right)\,,
\end{align}
and the radial momentum is
\begin{align}
p_{r}^{2}&=\frac{m_{2}^{2}}{B(r)}\bigg(\frac{\gamma^{2}}{A(r)}-1-\frac{(\gamma^{2}-1)b^{2}}{r^{2}} \nonumber \\
&+2c_{E}E^{2}(r,\gamma,b)+2c_{B}B^{2}(r,\gamma,b)\bigg)\,,
\end{align}
where for equatorial orbits in a Schwarzschild background
\begin{align}
E^2&=\frac{18J^{4}(Gm_{1})^2}{r^{10}}+\frac{18J^{2}(Gm_{1})^4}{r^{8}}+\frac{6(Gm_{1})^2}{r^{6}} \nonumber \\
B^{2}&=\frac{18J^{4}(Gm_{1})^2}{r^{10}}+\frac{18J^{2}(Gm_{1})^4}{r^{8}}\,.
\end{align}

In each case, we expand the radial momentum  perturbatively in both $G$ and the $\ell$, and evaluate the radial action integral
\begin{equation}
i_{r}=2\int_{b}^{\infty} \sqrt{p_{r}^{2}}\,
\end{equation}
where a hard cutoff is used as $r$ approaches $b$ and power-law divergences are discarded. Each term in the expansion is computed by the integral
\begin{equation}
\int_{b}^{\infty}dr\left(1-\frac{b^{2}}{r^{2}}\right)^{1/2-q}r^{-k}=b^{1-k}B(\tfrac{1}{2}(k-1),\tfrac{3}{2}-q)\,.
\end{equation}

\bibliographystyle{utphys-modified}

\bibliography{references}

\end{document}